\documentclass[aps,prl,reprint,superscriptaddress,floatfix]{revtex4-2}
\usepackage{siunitx}
\usepackage[caption=false]{subfig}
\usepackage{amsmath}
\usepackage{graphicx}
\usepackage{xcolor}
\usepackage{float}

\begin{document}

\title{Creating QED Photon Jets with Present-Day Lasers}
\author{Scott V. Luedtke}
\affiliation{Los Alamos National Laboratory, Los Alamos, New Mexico, 87545, USA}
\affiliation{University of Texas, Austin, Texas 78712, USA}
\author{Lin Yin}
\affiliation{Los Alamos National Laboratory, Los Alamos, New Mexico, 87545, USA}
\author{Lance A. Labun}
\author{Ou Z. Labun}
\affiliation{University of Texas, Austin, Texas 78712, USA}
\author{B. J. Albright}
\author{Robert F. Bird}
\author{W. D. Nystrom}
\affiliation{Los Alamos National Laboratory, Los Alamos, New Mexico, 87545, USA}
\author{Bj\"{o}rn Manuel Hegelich}
\affiliation{University of Texas, Austin, Texas 78712, USA}
\affiliation{Center for Relativistic Laser Science, Institute for Basic Science, Gwangju 61005, South Korea}
\affiliation{Department of Physics and Photon Science, Gwangju Institute of Science and Technology, Gwangju 61005, South Korea}

\date{\today}

\begin{abstract}
    Large-scale, relativistic particle-in-cell simulations with quantum electrodynamics (QED) models show that high energy (1$<E_\gamma\lesssim$~75~MeV) QED photon jets with a flux of $10^{12}$~sr$^{-1}$ can be created with present-day lasers and planar, unstructured targets.
    This process involves a self-forming channel in the target in response to a laser pulse focused tightly ($f$~number unity) onto the target surface.
    We show the self-formation of a channel to be robust to experimentally motivated variations in preplasma, angle of incidence, and laser stability, and present in simulations using historical shot data from the Texas Petawatt.
    We estimate that a detectable photon flux in the 10s of MeV range will require about 60~J in a 150~fs pulse.
\end{abstract}

\maketitle

The possibility of producing copious MeV-scale photons from short-pulse laser-matter interactions has attracted attention because of potential applications including laboratory astrophysics~\cite{bulanov2015problems}, radiation therapy~\cite{weeks1997compton}, and radiosurgery~\cite{girolami1996photon}.
Radiation dynamics are also important for other applications of laser-matter interactions, ranging from ion acceleration for cancer therapy~\cite{,bulanov2002oncological,fourkal2003intensity,malka2004practicability} to fast sources of x-rays for imaging~\cite{bloembergen1999nanosecond,rousse2001femtosecond,rousse2004production}, because they generally require more powerful lasers than are available today to produce high enough energies or fluxes of particles.
More powerful lasers broadly means more acceleration of charged particles, and therefore more energy lost to radiation.
Developing and validating accurate models of radiation is therefore crucial not only to applications of radiation, but applications of the radiating particles as well.

Ideally, models would be validated in specialized experiments using present-day lasers before being widely deployed as a predictive tool for experiments on future laser systems that are sure to have strong, quantum electrodynamic (QED), radiation effects.
Previous simulations have predicted large numbers of high-energy photons \cite{ridgers2012,ji2014energy,ji2014near,nerush2014gamma,zhu2015enhanced,stark2016enhanced,luedtke2018jet} that can help test QED models used in simulation codes.
However, none of these simulations have been realized experimentally because they rely on lasers more powerful than are currently operational, specialized micro-structured targets requiring precision pulse control, or both.
In this Letter, we present simulations of a novel and robust method for photon production involving a self-forming channel that may be achievable in laboratory experiments using operating lasers like the Texas Petawatt~\cite{martinez2012texas} or the CoReLS 4~PW laser~\cite{sung20174,yoon2019achieving}.
We discuss the robustness of channel formation to experimentally motivated perturbations and run simulations using historical shot data from the Texas Petawatt.

Short-pulse laser-plasma interactions are typically modeled using fully-kinetic particle-in-cell (PIC)~\cite{birdsall2004} simulations solving the Maxwell-Vlasov system of equations for the plasma distribution function, often with reduced spatial dimensions for computational reasons.
With emitted photon energies more than 100 times the electron rest mass, and a significant fraction of the emitting electron energy, classical models~\cite{vranic2016picrr} of radiation reaction are unsuitable.
We instead use a semi-classical model~\cite{Elkina:2010up} of the spin- and polarization-averaged emission rates in strong fields~\cite{Ritus:1985}.
For electrons in laser fields, high-energy photon emission rates depend on the Lorentz invariant $\chi=(e\hbar/m_e^3c^4)|F_{\mu\nu}(\vec{r})p^\nu|$, with $p^\nu$ the electron 4-momentum, and $F_{\mu\nu}(\vec{r})$ the electromagnetic field tensor.
The quantity $\chi$ can be viewed physically as the electron's acceleration in natural units and hence increasing the number and energy of emitted photons means increasing the acceleration experienced by the electrons.

\begin{figure*}
    \centering
	\includegraphics[width=0.27\textwidth]{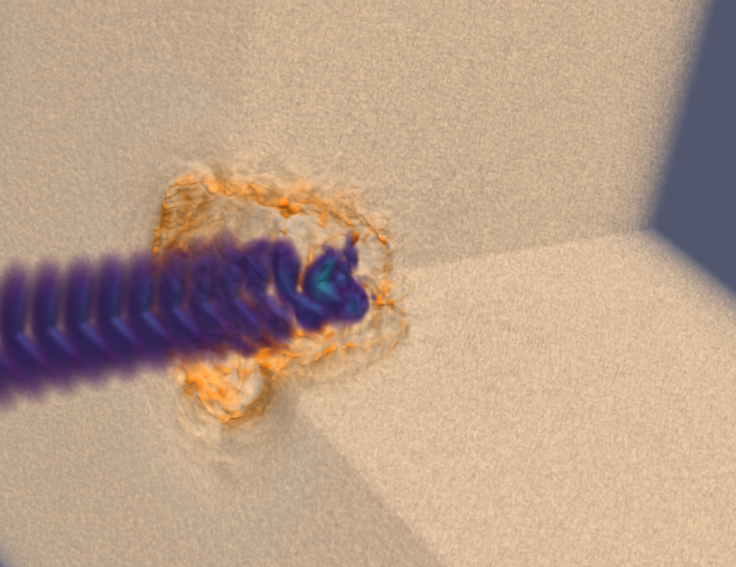}
	\includegraphics[width=0.27\textwidth]{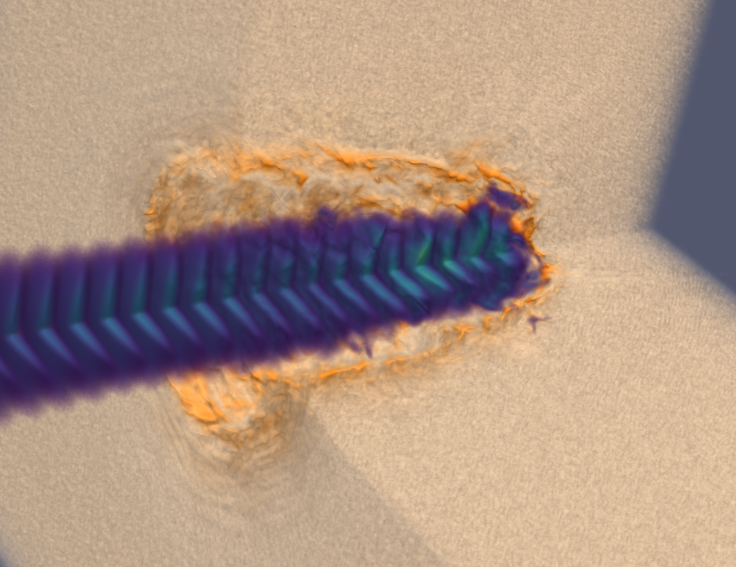}
	\includegraphics[width=0.27\textwidth]{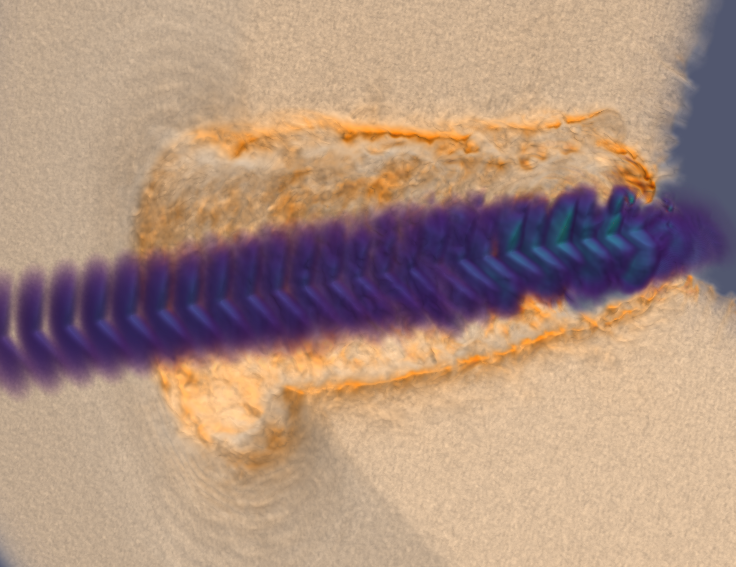}
	\includegraphics[width=0.145\textwidth]{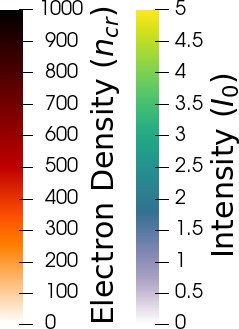}
    \caption{Visualization of the self-formation of a channel in a 3D simulation with a quadrant cut out showing enhanced fields in the channel.  The colorbars are clipped for visualization.  An animation is available in supplemental material.}
    \label{cool3D}
\end{figure*}

We first demonstrate the self-forming channel and enhanced intensity in a high-resolution, non-QED 3D simulation visualized in Fig.~\ref{cool3D}.
We use a focused Gaussian pulse with a wavelength $\lambda_l=2\pi c/\omega_l=$~\SI{1.058}{\micro\meter}, FWHM pulse duration in intensity of {150}~fs, peak intensity of $3.02 \times 10^{22}$ W/cm$^2$ (normalized laser amplitude $a_0=eE\lambda_/2\pi m_e c^2 = 157$), beam waist radius $w_0 = $~\SI{1.25}{\micro\meter} ($\sim f/1$ focusing), and hyperbolic secant temporal profile~\cite{McDonald:1998pu}.
The target is a fully ionized carbon plasma, \SI{10}{\micro\meter} thick, with initial electron density 90$n_{cr}$ (where $n_{cr}=\omega_l^2m_e/e^2$ is the critical density), electron temperature 10~keV and ion temperature 10~eV.

The simulation shows, as we explain below, the ion restorative force balances the ponderomotive expansion force on the electrons to form a channel in the target.
Very dense ($>400n_{cr}$) channel walls confine the laser pulse and enhance the intensity, which peaks at $9I_0$ and exceeds $4I_0$ in much of the channel.
By increasing the field strength in the channel, this configuration greatly enhances the probability of QED photon emission.
Though this simulation had no QED effects, we expect them to affect the channel-forming plasma dynamics very little, since only a few percent of the laser energy is converted into QED photons.
This simulation had considerably higher resolution than previous 3D simulations~\cite{ji2014energy,ji2014near,stark2016enhanced}, with 60 macroparticles per cell and a 9.2~nm cell size in each dimension, resulting in approximately two field points per electron skin depth and four Debye lengths per cell for our initially 10~keV-temperature target.
We used the highly efficient code \textsc{vpic}~\cite{bowers20080,bowers2008ultrahigh,bowers2009advances}.

In the remainder of this Letter, we utilize 2D simulations to analyze the self-formation of a channel and resulting $\gamma$-ray emission.
This allows us to run many simulations and study the effects of changing several parameters.
We use the \textsc{psc} \cite{psc}, which includes the QED model described above, with the same pulse and target parameters used in our 3D simulation, and with the laser polarization oriented out of the simulation plane~\cite{stark2017effects}.
To obtain physical units in our plots, we assume a thickness in the third dimension of $\sqrt{{\pi}/{2}}w_0$, which preserves the total energy in the laser pulse.

The number of photons $N_\gamma$ produced during a laser-matter interaction can be written as  
\begin{equation}
    N_\gamma = \int_{-\infty}^\infty\!dt\int_V^{} \!d^3\vec{r}\int \!d^3\vec{p}\ f_e(\vec{r}, \vec{p},t) \frac{dN_\gamma^{(1)}(\vec{r}, \vec{p}, t)}{dt},
\label{eq:N_gamma}
\end{equation}
where $f_e(\vec{r},\vec{p},t)$ is the electron phase space density and $dN_\gamma^{(1)}(\vec{r},\vec{p}, t)/dt$ is the photon emission rate from a single electron.
Maximizing $N_\gamma$ is an optimization problem of the density of the target and the intensity, duration, and spatial extent of the pulse.
In this work we do not use, for example, external electron beams, so momentum is not a significant optimization parameter.
Since $dN_\gamma^{(1)}(\vec{r}, \vec{p}, t)/dt$ is exponentially suppressed at low field strength, we expect to sacrifice laser duration and spatial extent for maximum intensity, using as short a pulse and as tight a focus as possible.
The target optimization is complex, since it can affect the laser intensity and spatial extent via modified plasma conditions.

For the laser and target parameters examined in this Letter, light pressure far exceeds plasma pressure throughout the laser-plasma interaction. 
Thus the formation of a channel through which the laser propagates is governed by the ponderomotive force from the laser, which exerts a force on a fluid element proportional to the gradient of the intensity. 
Electrons inside the laser spot are displaced forward in the laser field. 
This sets up a charge-separation electrostatic field. 
The balance of the ponderomotive force on the electrons and ion restoring force sets the hole-boring speed along the direction of laser propagation.
A similar process occurs in the direction transverse to the laser propagation, causing the channel to expand transversely, though the fields, and thus the ponderomotive force, drop dramatically about one beam waist, $w_0$, from the laser axis.

\begin{figure}
    \centering
	\includegraphics[width=\columnwidth]{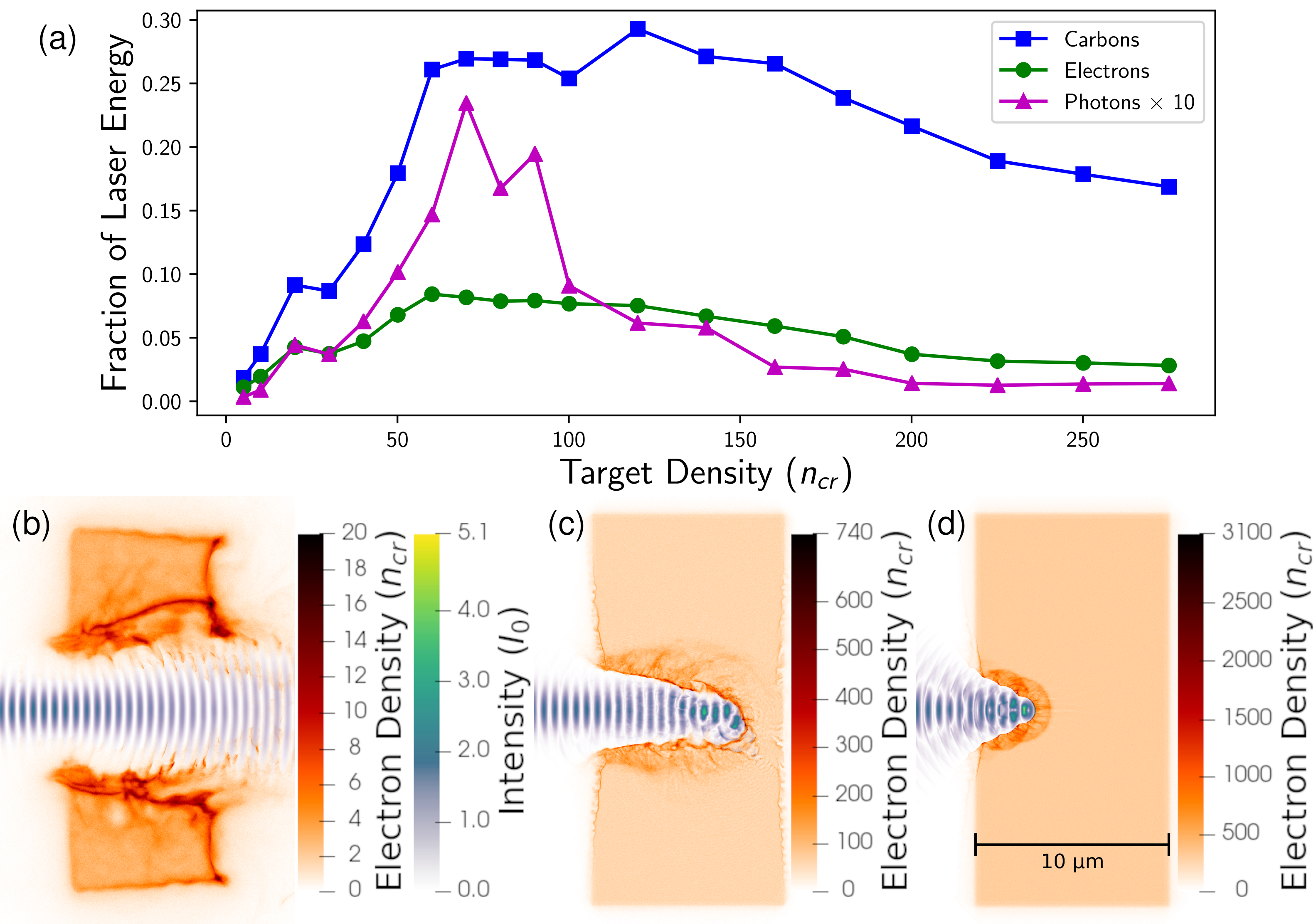}
    \caption[Small multiples]{(a) Fraction of laser energy transferred to each particle species at the end of the simulation as a function of target density (fixed \SI{10}{\micro\meter} target thickness).  Bottom: Electron density and laser E-field at the same time step, near when the peak of the pulse arrives at the target, for $3n_{cr}$ (b), $60n_{cr}$ (c), and $300n_{cr}$ (d).}
    \label{fig:densitysnaps}
\end{figure}

From Eq.~\ref{eq:N_gamma}, photon production increases with target density, laser intensity, and laser-plasma interaction volume.
However, during the channel formation, increasing density and intensity works against the requirement for large volume.
In the low target density limit, the laser ponderomotive force forms a wide channel with low electron density before the arrival of the peak intensity (see Fig.~\ref{fig:densitysnaps}), leading to low photon production.
In the opposite limit of a highly over-dense target, the transverse expansion and hole boring of the channel is limited.
The electron density at the edge of the channel is high from ponderomotive compression of the target, which reflects the pulse back on itself---increasing intensity, but only in a small volume from which electrons have already been evacuated, leading to low net photon production. 

\begin{figure}
    {\includegraphics[clip, trim=1.5in .1in .3in .04in, width=0.6\columnwidth]{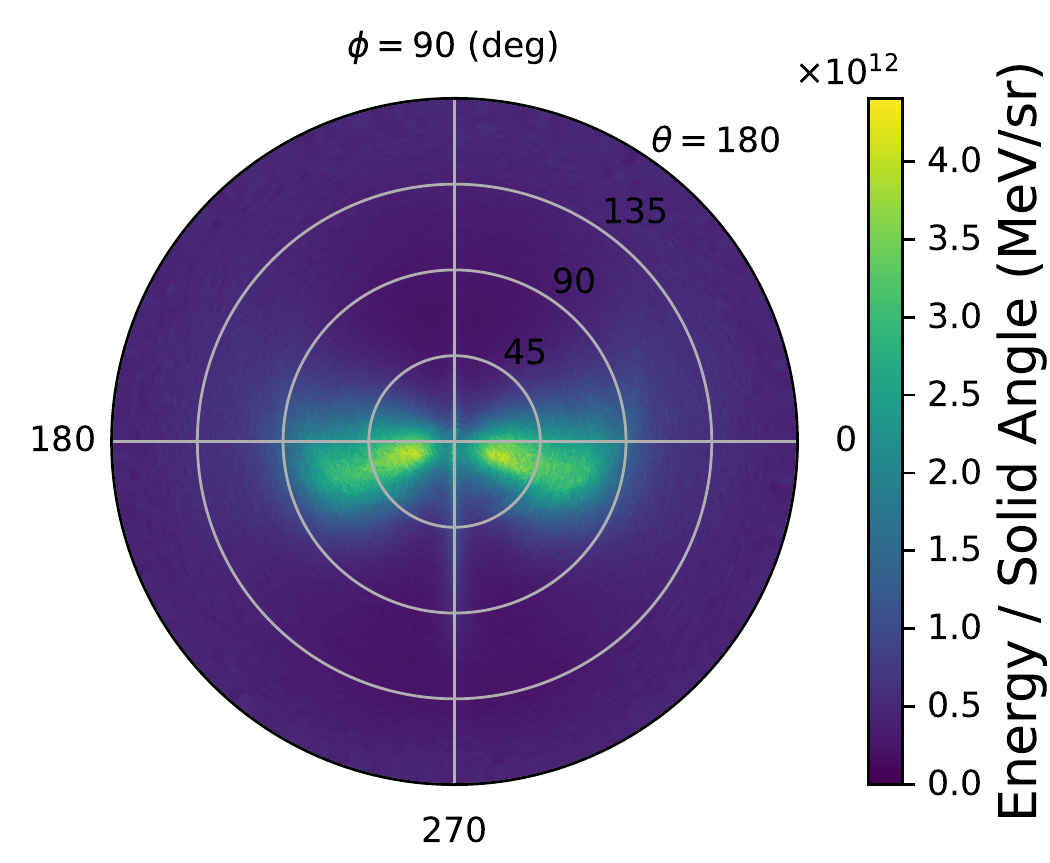}}\\
    \caption[Small multiples]{Angular energy flux of high-energy photons ($E_\gamma>1$\,MeV) for a simulation of a \SI{10}{\micro\meter} $n_e=60n_{cr}$ slab target. The laser propagates in the $(0^\circ,0^\circ)$ direction and is polarized along the $\phi=0^\circ$ plane.}
    \label{fig:phot_edep}
\end{figure}

At the optimal target density, approximately 50--100$n_{cr}$---readily available carbon foam targets---the ion restorative force balances the ponderomotive force to form a channel of radius $\sim w_0$, enhancing the intensity of the pulse as it interacts with electrons in the front of the channel, maximizing $dN_\gamma^{(1)}(\vec{r}, t)/dt$, while maintaining moderate density and having larger spatial extent than just the focal spot.
Electrons towards the front of the channel see a sudden increase in field strength as the channel-constrained pulse arrives and begin relativistic oscillatory motion, similar to that of a free electron in a plane wave.
These electrons exhibit much of the highest energy photon emission and the two-jet pattern seen in Fig.~\ref{fig:phot_edep}.
The two-jet pattern is similar to synchrotron radiation from an undulator, and is approximately what is expected from a single electron accelerating in a plane wave.

Previous works have predicted high photon fluxes in two jets, like described above, but have not been realized in experiment.

Refs.~\cite{ji2014energy,ji2014near} observe a two-jet pattern for linearly polarized pulses with intensities higher than so far realized in lab, and show this changes to an azimuthally symmetric ring for circularly-polarized pulses.
Ref.~\cite{ji2014energy} further showed that the jets merge in the forward direction at about $a_0=200$ for a $32n_{cr}$ hydrogen target and $a_0=1000$ for $300n_{cr}$ carbon.

In Ref.~\cite{zhu2015enhanced}, a hydrogen-filled gold cone is used to further focus a laser pulse to higher intensities.
They rely on the radiation-reaction trapping effect~\cite{ji2014radiation} to keep electrons under the enhanced-intensity pulse.
The intensity $a_0=180$ has not been demonstrated in the lab with their large beam waist $w_0 = $ \SI{5}{\micro\meter}.
Despite a linearly polarized pulse, the photon distribution in these simulations displays a single peak around the laser axis, suggesting that, as in Ref.\cite{ji2014energy}, the two jets have merged because of the enhanced intensity provided in the cone.
Experiments with lower total pulse energy but similar peak intensities could show significant emission, but such a complex target has not been built.

Simulations in~\cite{stark2016enhanced} use a structured target with a low-density ($\sim 10n_{cr}$) channel about the width of the laser focus surrounded by a high-density ($\sim 100n_{cr}$) enclosure.
The laser parameters are somewhat optimistic for today's laser systems with $a_0=190$, $w_0=$\SI{1.1}{\micro\meter}, and FWHM = 100~fs.
The high-density enclosure confines the pulse to the channel and maintains the intensity at the focus as the pulse travels through the channel, an engineered analog to the self-organizing dynamics in our simulations with unstructured targets.
However, the typical pointing stability of such tightly focused lasers is on the order of a few tens of microrad, i.e., multiple focus radii, meaning that obtaining statistically significant data from a micron-scale feature will require many more shots than typically available in experimental campaigns on petawatt systems, and initial experiments have proven to be challenging.
Other proposals~\cite{vranic2018extremely,wang2020power} to use structured targets face similar issues.

A feature of our simulations that was not remarked upon in previous work is that the plasma dynamics vary stochastically.
Changing the random seed in our simulations---which changes the \emph{microscopic} state of the initial plasma, i.e., the precise position and momentum of individual particles but not mean density or energy---significantly affects the \emph{macroscopic} dynamics of the self-forming channel.
For example, the formation of the channel can deviate from the laser axis by up to \SI{20}{\degree}, and the resulting photon jets follow this deviation.
We explore these shot-to-shot variations and resolution requirements in more detail in another work~\cite{luedtke2020simulations}.
Additional shot-to-shot variations can result from variances in experimental parameters, which has been neglected in previous simulation work, but we will address below.

A self-forming channel preserves much of the QED emission from previous work while obviating many experimental challenges.
We have further investigated the robustness of self-forming channels by testing conditions more representative of laboratory experiments and sensitivity to several experimental parameters.
We give a brief overview of these simulations here and present the results and in-depth analysis in a longer work~\cite{luedtke2020simulations}.

Most experiments orient the target at an angle to the laser propagation direction in order to suppress retro-reflection, which can damage components in the laser chain.
Testing this in simulation, the jet axes are offset from target normal by the laser incidence angle.
When shot at an angle, the pulse effectively sees a thicker target, and these simulations produce more photons.
Scanning target thickness indicates that about \SI{25}{\micro\meter} is the limit of the channel depth, but with no penalty for thicker targets, which may be easier to deploy.

The plasma density profile at the start of a simulation may be unknown because a prepulse or other deviations from ideal assumptions about the pulse profile can cause pre-expansion of the plasma, which impacts the laser-plasma dynamics and final particle energies \cite{batani2010prepulses,sentoku2017}.
Measuring this profile has proven to be an extraordinary experimental challenge.
Estimating a pre-expanded plasma state in simulation can nevertheless give insight into how a preplasma might affect experiments.
We model the preplasma as a Gaussian envelope in front of the target with a standard deviation of \SI{5}{\micro\meter}.
The preplasma results in about 40\% more energy converted into photons, a rise from 1.8\% to 2.5\% total conversion efficiency averaged over five simulations.
In the lower density preplasma, the less-focused and early parts of the pulse form a focusing channel that reliably produces a tighter channel and higher energy photons.
Simulations with both a preplasma and angle-incidence behaved similarly to simulations with just a preplasma, but with the channel forming in the laser direction.

Since the emission probability for high-energy photons is exponentially suppressed for small $\chi$, photon yield and energy should be sensitive to the laser intensity.
We find that to be the case, with a doubling of the pulse energy resulting in at least an order of magnitude increase in the photon flux.
Our scan indicates that about 60~J are required for measurable flux in the 10s of MeV range, considering that gradient-magnetic gamma spectrometers~\cite{tiwari2019gradient} require roughly $10^8$~photons/(sr~MeV) for a detectable signal.
Shorter pulses that are less energetic but more intense may have different requirements for significant photon production.

\begin{figure}[htp]
    \centering
	\includegraphics[width=0.9\columnwidth]{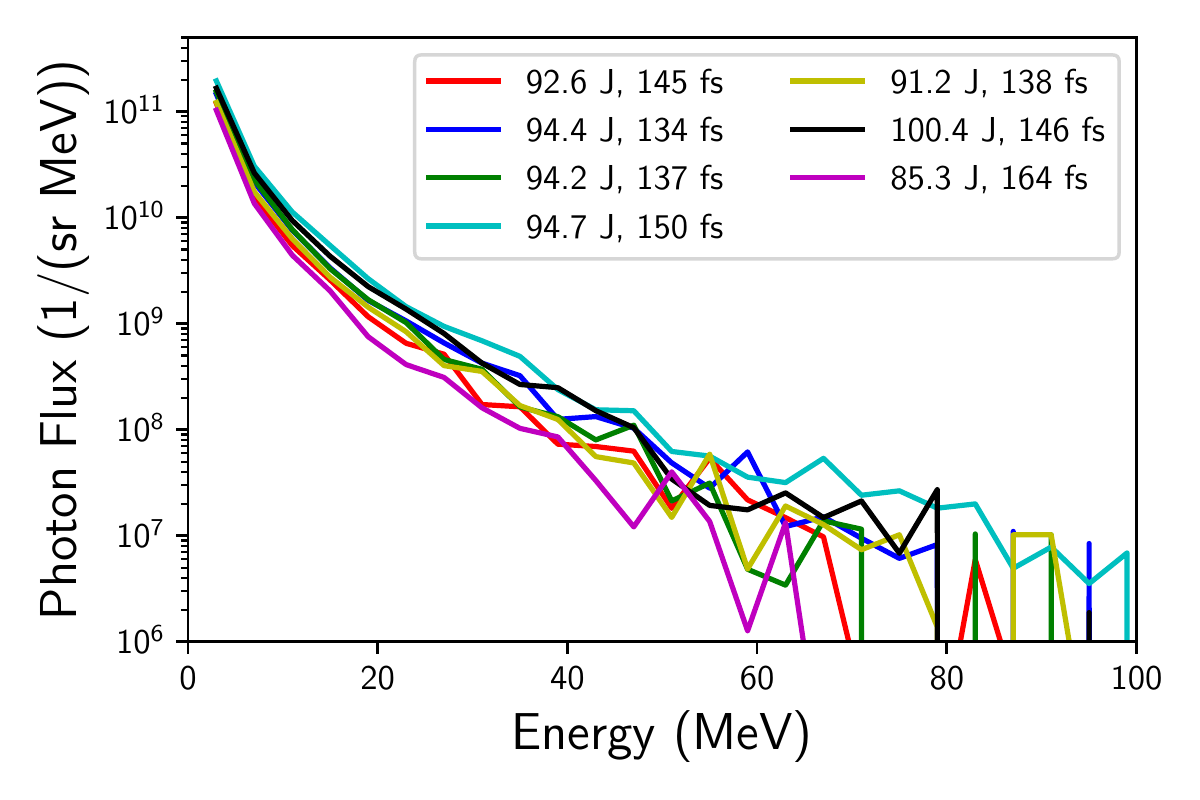}
   \caption[Small multiples]{Pinhole photon flux at $(\phi, \theta)=(0^\circ,20^\circ)$ for simulations using real shot data from the Texas Petawatt demonstrating the variation expected when considering laser instabilities in energy and pulse duration. \label{fig:TPW_20_0}}
\end{figure}

Given the high sensitivity of photon production to laser energy, a natural question is how laser stability affects the reproducibility of experiments.
Laser parameters at full power on-target are difficult or impossible to measure, with the most common measurements coming from a pick-off mirror after pulse compression, but before focusing.
To explore this, we obtained 50 shot reports from the Texas Petawatt.
We ran simulations for the first five shots and the highest and lowest energy of the 50 total shot reports using a 60$n_{cr}$, \SI{10}{\micro\meter} carbon target with a \SI{5}{\micro\meter} preplasma.
Assuming $f/1$ focusing and a hyperbolic secant temporal pulse profile, the energy and duration from each shot report defines the pulse.
The energies across the seven reports varied from 85.3--100.4~J, pulse duration from 134--164~fs, and (calculated) intensity from 1.86--$2.54\times 10^{22}$~W/cm$^2$.
The resulting photon flux in a pinhole near the center of the average jet is shown in Fig.~\ref{fig:TPW_20_0}.
The shot to shot variation in the photon flux in the five Petawatt simulations is about twice that for simulations with the same pulse.
For example, at 30~MeV, the standard deviation of the flux relative to the average flux for five simulations with the same pulse is 23\%, compared to 41\% for the five Petawatt simulations.
Laser instabilities in energy and pulse duration should not be detrimental to an experimental campaign to detect photon jets, but will increase the number of shots required for good statistics.

Because both laser fluctuations and above-mentioned stochastic plasma dynamics yield photon spectra that vary by $\sim1$ order of magnitude from shot to shot, measuring the photon spectrum observed on a single sight-line will provide limited evidence for the channeling dynamics seen or correctness of the photon emission model employed in the simulations.
Using multiple detectors in different locations across repeated experiments, or even different measurements of the photon distribution \cite{jetsInPrep}, will give much stronger evidence for the process we describe.

Even using real shot data, we fail to account for many experimental conditions.
For example, the laser wavefront is likely distorted.
The laser temporal profile does not match any profile used in simulations, and can have large prepulses nanoseconds before our 1.5~ps simulation starts.
A real laser pulse will usually deviate from an ideal shape at a level of $10^{-2}$--$10^{-5}$ in intensity and thus sit on a pedestal of laser intensity in time and/or space that can modify the plasma conditions to an extent where the interaction with the peak of the pulse is changed.
These practical technical issues of spatio-temporal couplings tend to get more severe the shorter and more tightly focused the pulse is.
Another example is that our simulations ignore collisions and ionization.
In laboratory experiments involving high intensity laser-target interaction, collisions would allow absorption of laser energy by the target plasma via processes such as inverse bremsstrahlung.
Together with $\vec{J}\times \vec{B}$ laser heating, these processes would quickly increase the target temperature to that used initially in our simulations.
Although it is difficult to obtain measurements of this quick temperature increase, evidence of the target expansion as a consequence of the temperature increase at early times before the target becomes transparent to the laser is found in Trident experiments from the reflected light diagnostics as indicated by the blue shift in the spectra~\cite{palaniyappan2012dynamics}.
Further, we ignore bremsstrahlung radiation because a previous study~\cite{Wan:2017jhn} showed that it is not significant at these intensities for aluminum, and is therefore less important for our lower-Z carbon targets.
Lastly, we also ignore anything that happens outside our simulation volume ($\sim$\SI{30}{\micro\meter\cubed}), most notably bremsstrahlung radiation from other parts of the apparatus.
Distinguishing bremsstrahlung photons from strong-field QED photons will be essential to testing the QED models currently in PIC codes, but should be possible contrasting the distinctive two-jet pattern of QED photons with the radially symmetric ring expected from bremsstrahlung.

In conclusion, we have predicted a new way to construct a channel in an intense laser-plasma experiment that should be testable in the lab with today's lasers.
The channel results in a distinctive pattern of two jets of high-energy photons with measurable fluxes.
We have considered many ways in which real experiments differ from most simulations and concluded that the self-formation of a channel is robust to these perturbations.
The observation of these jets (or their absence) would be a big step towards validating QED models used in PIC codes and give confidence in their use to design future experiments and engineer applications.

\begin{acknowledgments}
Work performed under the auspices of the University of Texas at Austin, the U.S. DOE by Triad National Security, LLC, and Los Alamos National Laboratory .
This work was supported by the Air Force Office of Scientific Research (FA9550-14-1-0045) and the LANL ASC and Experimental Sciences programs.
High performance computing resources were provided by the Texas Advanced Computing Center and LANL Institutional Computing facilities.
This work used the Extreme Science and Engineering Discovery Environment (XSEDE), which is supported by National Science Foundation grant number ACI-1548562.
Thanks to R.~A.~Roycroft for providing Texas Petawatt shot reports and G.~Tiwari for discussions of gradient-magnetic spectrometers.
\end{acknowledgments}

\bibliography{CreatingQEDJets}

\end{document}